\documentclass[twocolumn,secnumarabic,amssymb, nobibnotes, aps, prm,superscriptaddress,onecolumn,preprint]{revtex4-2}

\usepackage{graphicx}
\usepackage{xcolor}
\usepackage[english]{babel} 
\usepackage[separate-uncertainty = true]{siunitx}

\usepackage{natbib}
\setcitestyle{square, comma, sort}
\begin{document}

\title{Preparation and STM study of clean Nb(111) surfaces}%

\begin{abstract}
Niobium with its highest transition temperature among all elemental superconductors has become a favorable substrate for realizing well-defined low-dimensional magnet-superconductor hybrid systems exhibiting novel types of exotic electronic states such as Majorana zero-energy modes. While a preparation procedure for obtaining atomically clean Nb(110) substrates has previously been reported, a suitable preparation method for clean Nb(111) surfaces is still lacking. Here, we report a recipe for cleaning Nb(111) surfaces based on an atomic hydrogen treatment followed by short flashes to elevated temperatures. The atomic surface structure of clean non-reconstructed Nb(111) is investigated by high-resolution scanning tunneling microscopy (STM), as well as a surface reconstruction with a reduced atom density compared to the (111) plane of a bcc crystal resulting from a surface premelting at high annealing temperatures.
\end{abstract}

\author{Julia J. Goedecke}%
\email{julia.goedecke@physnet.uni-hamburg.de}
\affiliation{Department of Physics, University of Hamburg, Hamburg, Germany}
\author{Maciej Bazarnik}%
\affiliation{Department of Physics, University of Hamburg, Hamburg, Germany}
\affiliation{Institute of Physics, Westfälische Wilhelms-Universität, Münster, Germany}
\affiliation{Institute of Physics, Poznan University of Technology, Piotrowo 3, 60-965 Poznan, Poland}
\author{Roland Wiesendanger}%
\email{wiesendanger@physnet.uni-hamburg.de}
\affiliation{Department of Physics, University of Hamburg, Hamburg, Germany}
\date{February 2023}%
\maketitle

\section{Introduction}
\label{sec:intro}
The preparation of atomically well defined low-dimensional magnet-superconductor hybrid (MSH) systems has been identified as a crucial step towards realizing topologically non-trivial electronic states, including Majorana zero-energy modes, which offer great potential for fault-tolerant topological quantum computing \cite{Kitaev2003, Nayak2008, Sarma2015}. Indeed, evidences for Majorana states have been found in 1D MSH structures \cite{Yazdani2014, Kim2018, Schneider2022} as well as 2D MSH systems \cite{Menard2017, Palacio2019}. In these studies, Pb(110), Re(0001) and Nb(110) have been used as elemental superconducting substrates. While (110) oriented surfaces with their two-fold symmetry are most adequate for realizing 1D atomic wires by either self-assembly or scanning tunneling microscope (STM) based single-atom manipulation processes, the (0001) or (111) surfaces with their threefold-symmetry are interesting for realizing 2D MSH structures, in particular those which potentially host magnetic skyrmions within the monoatomic magnetic layer. Indeed, magnetic skyrmions on an s-wave superconductor can host Majorana bound states \cite{YangG2016} and offer great potential for their manipulation as required for braiding operations \cite{Pershoguba2016, YangG2016, Mohanta2021, Mascot2021, Garnier2019}.\\
The Nb(111) surface would be an ideal choice for realizing promising 2D MSH systems due to the high transition temperature (above 9 K) of Nb between the normal conducting and superconducting state as well as its threefold symmetry, allowing for the growth of ultrathin magnetic layers with potential skyrmionic states. However, niobium surfaces are difficult to clean from oxygen and other contamination species. For Nb(110), the oxygen contamination problem has been solved by flashing the niobium crystal to very high temperatures, as described by Odobesko et al.~\cite{Odobesko2019}. However, this method cannot be used for Nb(111) since this surface exhibits a rather open atomic structure, in contrast to Nb(110). If the Nb(111) surface is heated to very high temperatures, i.e.~$T>~$\SI{2013}{K}, as required for removing the oxygen from the crystal surface, a rearrangement of the surface atoms occurs, leading to surface facetting. 
In this work, a recipe for the preparation of an atomically clean surface of a Nb(111) single crystal will be described. The crucial step towards the achievement of this goal is demonstrated to be an atomic hydrogen surface treatment. It will be shown that a well ordered and atomically clean Nb(111) surface can be obtained as a promising starting point to realize well defined 2D MSH structures. Atomic-resolution STM data reveals the atomic surface structure of clean non-reconstructed Nb(111). On the other hand, a reduced atom density compared to the truncated (111) plane of a bcc crystal can be explained by a premelting of the surface.

\section{Methods}
\label{sec:methods}
The experiments were performed in a home-built versatile variable-temperature (25 K – 300 K) scanning tunneling microscope (VT-STM) system with home-built ultra-high vacuum (UHV) chambers \cite{Kuck2008}.
STM data were recorded with a tungsten tip at constant tunnel current $I$ with a bias voltage $V$ applied to the sample, for which the corresponding values are given in the figure captions. A Nb(111) crystal was initially cleaned by repeated cycles of argon ion sputtering and annealing using an electron beam heater. For the first cycles, a sputtering time of 60 min was used to initially remove surface oxide layers, with an annealing time of 10 min.
After twelve cycles, the sputtering time was reduced to 30 min, whereas the annealing time remained the same. However, the annealing temperature was increased step-wise as described below. After each cycle, the Nb crystal was transferred to the STM stage and investigated with respect to topographic features and atomic-scale surface structures. After noticing that Nb(111) could not be prepared properly by this method, an alternative chemical approach was chosen to achieve a clean and well ordered Nb(111) surface. For this purpose, a hydrogen cracker was used to chemically reduce niobium oxide with atomic hydrogen.
The Nb(111) sample was exposed to atomic hydrogen at a pressure of $p~=~10^{-7}~\mathrm{mbar}$ for $t$~=~\SI{16}{h}, with additional annealing at $T~\approx$~\SI{473}{K} and short flashes up to $T~\approx$~\SI{1273}{K} afterwards. The corresponding influences on the surface topography and atomic-scale surface structure were checked by high-resolution STM measurements.

\section{Results und Discussion}
\subsection{Attempts of Cleaning Nb(111) by Sputter-anneal Cycles}
\begin{figure}[htbp]
	\includegraphics[scale=0.8]{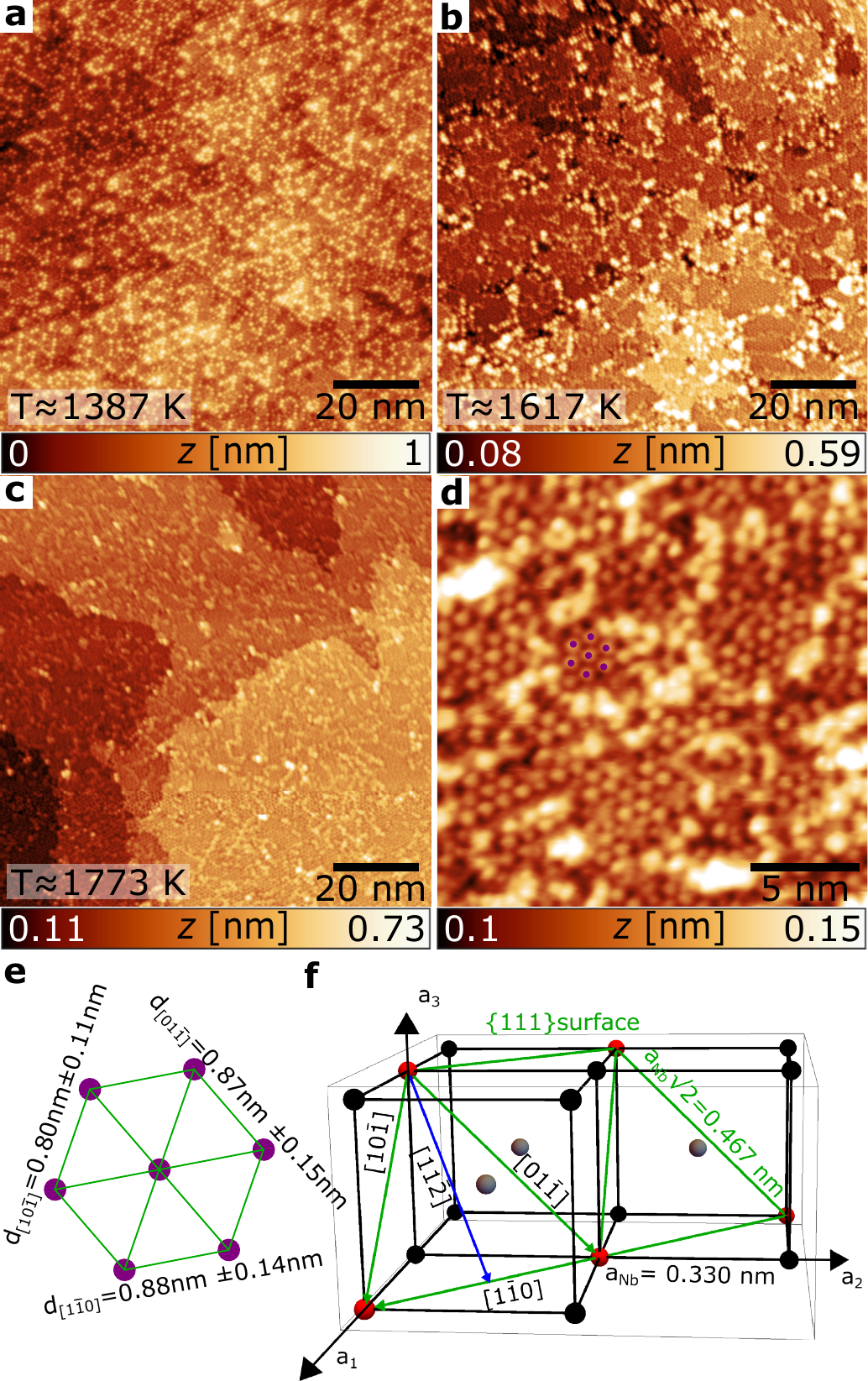}
	\caption{\label{fig:Nb111_1}
		STM images of the Nb(111) surface after different annealing treatments and atomic-scale STM data of the Nb(111) surface. (\textbf{a}-\textbf{c}) STM data showing how the Nb(111) surface gradually forms smooth terraces separated by well defined monoatomic steps as the annealing temperature is increased to (\textbf{a}) $T~\approx$~\SI{1387}{K}, (\textbf{b}) $T~\approx$~\SI{1617}{K} and (\textbf{c}) $T~\approx$~\SI{1733}{K}. 
		(\textbf{d}) Close-up view of the Nb(111) surface after annealing at $T~\approx$~\SI{1773}{K} showing small clusters (bright protrusions) on top of a hexagonal superlattice structure. (\textbf{e}) Schematic sketch of the underlying hexagonal superlattice as appearing in (\textbf{d}) with measured distances. The bcc crystal structure of Nb is shown for comparison (\textbf{f}).
		The measurement parameters are (a): $V$~=~\SI{500}{mV}, $I$~=~\SI{150}{pA}, (b): $V$~=~\SI{500}{mV}, $I$~=~\SI{120}{pA}, (c): $V$~=~\SI{500}{mV}, $I$~=~\SI{50}{pA}, (d) $V$~=~\SI{500}{mV}, $I$~=~\SI{50}{pA}, with $T~\approx$~\SI{25}{K}.}
\end{figure}

\begin{figure}	
	\includegraphics[scale=0.8]{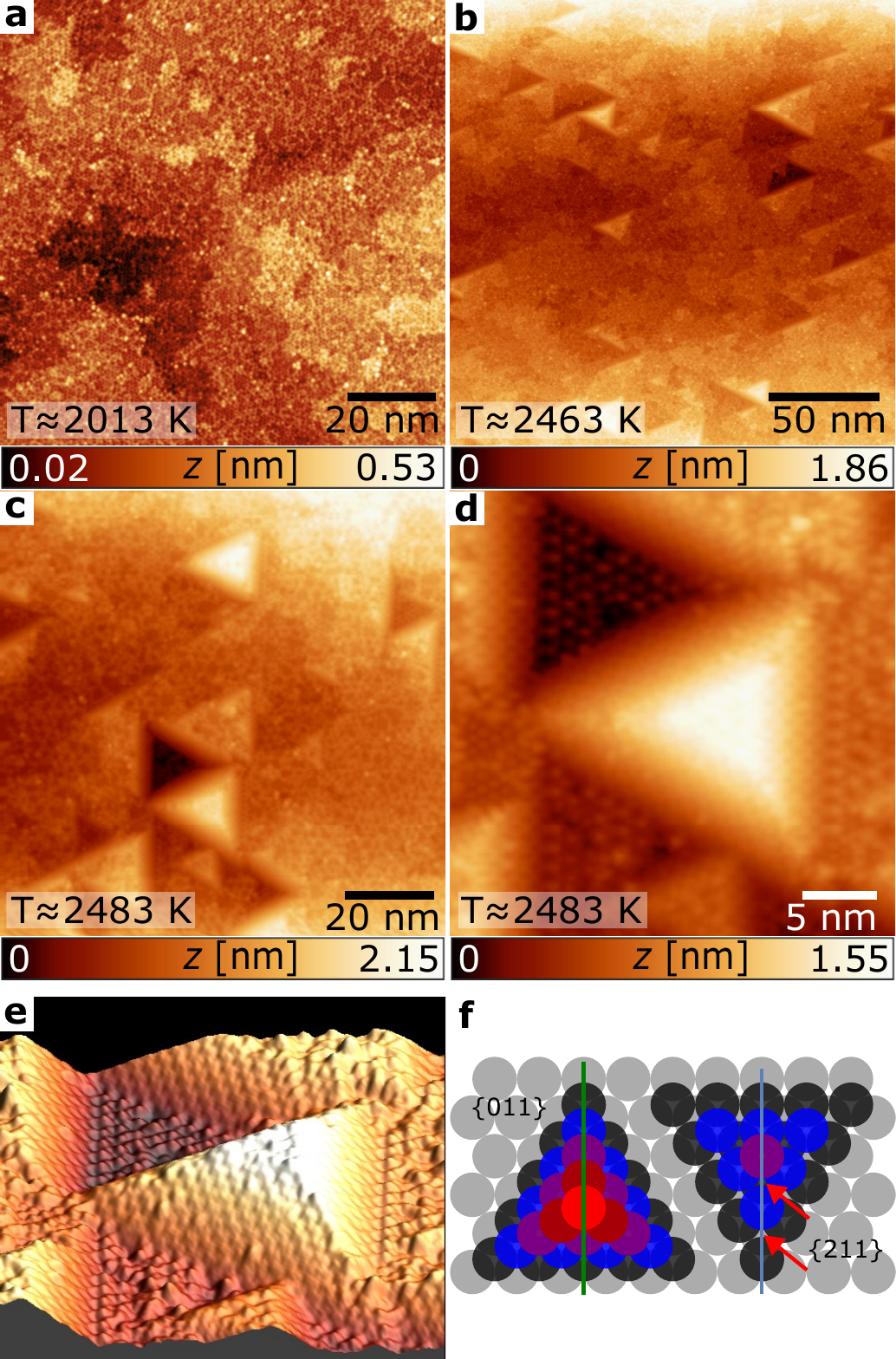}
	\caption{\label{fig:Nb111_2}
		STM images of the Nb(111) surface after annealing at higher temperatures and close-up view of a pyramid. (\textbf{a}) STM data of the Nb(111) surface after annealing at $T$~=~\SI{2013}{K}. No extension of the average terrace size is visible. Instead, if the crystal is heated further to $T$~=~\SI{2463}{K}, pyramid-like hillocks and depressions form on the surface (\textbf{b}) which continue to grow in height and depth when the annealing temperature is even further increased to $T$~=~\SI{2483}{K} (\textbf{c}). Close-up view of a pyramid. After annealing to $T~\approx$~\SI{2483}{K}, the Nb surface most likely exhibits $\left\lbrace 211\right\rbrace$ pyramidal structures (\textbf{d}). A 3D representation of the surface morphology is presented in (\textbf{e}).
		In (\textbf{f}) a schematic sketch (inspired by [23]) is showing the difference between for a $\left\lbrace 011\right\rbrace$- and $\left\lbrace 211\right\rbrace$-pyramid structure where the blue vertical line indicates the expected orientation of the line profile shown in the Supplementary Figure S1 panel (\textbf{a}) taken over the pyramid structure in the STM image of the Supplementary Figure S1 panel (\textbf{b}).
		STM measurement parameters: (\textbf{a}) $V$~=~\SI{500}{mV}, $I$~=~\SI{50}{pA}, (\textbf{b}) $V$~=~\SI{-2}{V}, $I$~=~\SI{50}{pA}, (\textbf{c}) $V$~=~\SI{2}{V} and (\textbf{d}) $V$~=~\SI{-750}{V}, $I$~=~\SI{500}{pA} with $T$~=~\SI{25}{K}.}
\end{figure}

First, the Nb(111) single crystal has been prepared by cycles of Ar ion sputtering and annealing at different temperatures. Fig.~\ref{fig:Nb111_1} \textbf{a} shows a STM image of the Nb(111) surface after it has been annealed to $T~\approx$~\SI{1387}{K} for 10 min. The surface exhibits nano-scale roughness, and no well defined atomic terraces have formed yet. Accordingly, a temperature around \SI{1400}{K} is not sufficient to heal the surface from sputtering after 10 minutes of annealing. Increasing the annealing temperature to $T~\approx$~\SI{1617}{K} (Fig.~\ref{fig:Nb111_1} \textbf{b}) leads to the formation of terraces, however, with some nano-scale holes and clusters on top. The STM image in Fig.~\ref{fig:Nb111_1} \textbf{c} shows the Nb(111) surface after annealing to $T~\approx$~\SI{1773}{K}. Compared to Fig.~\ref{fig:Nb111_1} \textbf{b}, the terraces are larger and mostly closed. In a smaller scale STM image, as presented in Fig.~\ref{fig:Nb111_1} \textbf{d}, a  hexagonal ($2 \times 2$) superlattice can clearly be identified. The distances of the hexagonal superlattice, as measured in the different directions (see schematic sketch in Fig.~\ref{fig:Nb111_1} \textbf{e}), are approximately twice the atomic distances of the hexagonal Nb(111) lattice (Fig.~\ref{fig:Nb111_1} \textbf{f}). 
These findings are in agreement with those of Coupeau et al.~\cite{Coupeau2015} who suggested that a clean Nb(111) surface is associated with a ($2 \times 2$) reconstruction, characterized by a lattice constant twice as large as for a truncated (111) plane of a bcc Nb crystal and therefore with an atom density reduced by a factor of 4. Furthermore, elongated clusters (bright features) on the surface can be seen in the STM image of Fig.~\ref{fig:Nb111_1} \textbf{d}.
It is most likely that these clusters are due to niobium oxide species still present on the surface. Since  $\mathrm{Nb}_2\mathrm{O}_5$ is already reduced at $T~\approx$~\SI{423}{K} \cite{Qing2004, Zhussupbekov2020} and $\mathrm{Nb}\mathrm{O}_2$ is reduced at $T~\approx$~\SI{613}{K} \cite{Delheusy2008,Zhussupbekov2020}, it can be assumed that these clusters are related to the presence of $\mathrm{NbO}$. Based on our own experience with the preparation of Nb(110) and reports from the literature \cite{Haas1966, Odobesko2019, Razinkin2008, Arfaoui2004}, $\mathrm{NbO}$ is a very stable oxide, which in the case of Nb(110) evaporates from the crystal surface only at temperatures above $T~\approx$~\SI{2730}{K}. Therefore, analogous to Nb(110), the same procedure was followed for Nb(111), i.e.~heating the crystal to higher temperatures. 
In Fig.~\ref{fig:Nb111_2} \textbf{a} an STM image of the crystal surface, which was annealed for 10 min at about $T~\approx$~\SI{2013}{K}, is presented. Interestingly, no further enlargement of the average terrace size was found in comparison with the STM image of Fig.~\ref{fig:Nb111_1} \textbf{c}, where the Nb(111) crystal has been annealed up to \SI{1773}{K}. On the other hand, the clusters, which are most likely related to remaining NbO on the surface, are still present.
This observation suggests that the annealing treatment at the higher temperature of around \SI{2000}{K} leads to oxygen diffusion from the interior of the crystal to its surface, while this temperature is not high enough to evaporate oxygen from the surface. 
After annealing the Nb(111) crystal at $T~\approx$~\SI{2463}{K}, STM images revealed the formation of triangular-shaped islands and vacancy islands (see Fig.~\ref{fig:Nb111_2} \textbf{b}). Then, after further annealing up to $T~\approx$~\SI{2483}{K}, these progressively transformed into pyramid structures (Fig.~\ref{fig:Nb111_2} \textbf{c}), as can most clearly be seen in smaller scale STM images (Fig.~\ref{fig:Nb111_2} \textbf{d} and \textbf{e}). This observation is typical for a surface where facetting occurs: An initially planar surface transforms into a ’hill-and-valley’ structure \cite{Madey1999}. 
Possible pyramid structures on a bcc(111) surface are the $\left\lbrace 011\right\rbrace$ and the $\left\lbrace 211\right\rbrace$-pyramid \cite{Madey1999}, as shown schematically in Fig.~\ref{fig:Nb111_2} \textbf{f}.
Madey et al.~\cite{Madey1999} shows that the $\left\lbrace 211\right\rbrace$ pyramid structure has a lower total surface energy than the $\left\lbrace 011\right\rbrace$ structure. In addition, step edges, indicated in the Supplementary Figure S2 \textbf{a} by the red arrows, could be found, which support the hypothesis of a  $\left\lbrace 211\right\rbrace$ pyramid structure.\\
In conclusion, an atomically flat and well ordered Nb(111) surface free of oxide and other surface contaminations could not be obtained solely based on cycles of ion sputtering and annealing.\\

\subsection{Niobium (111) Cleaning Treatment With Atomic Hydrogen}
\begin{figure}	
	\includegraphics[scale=0.8]{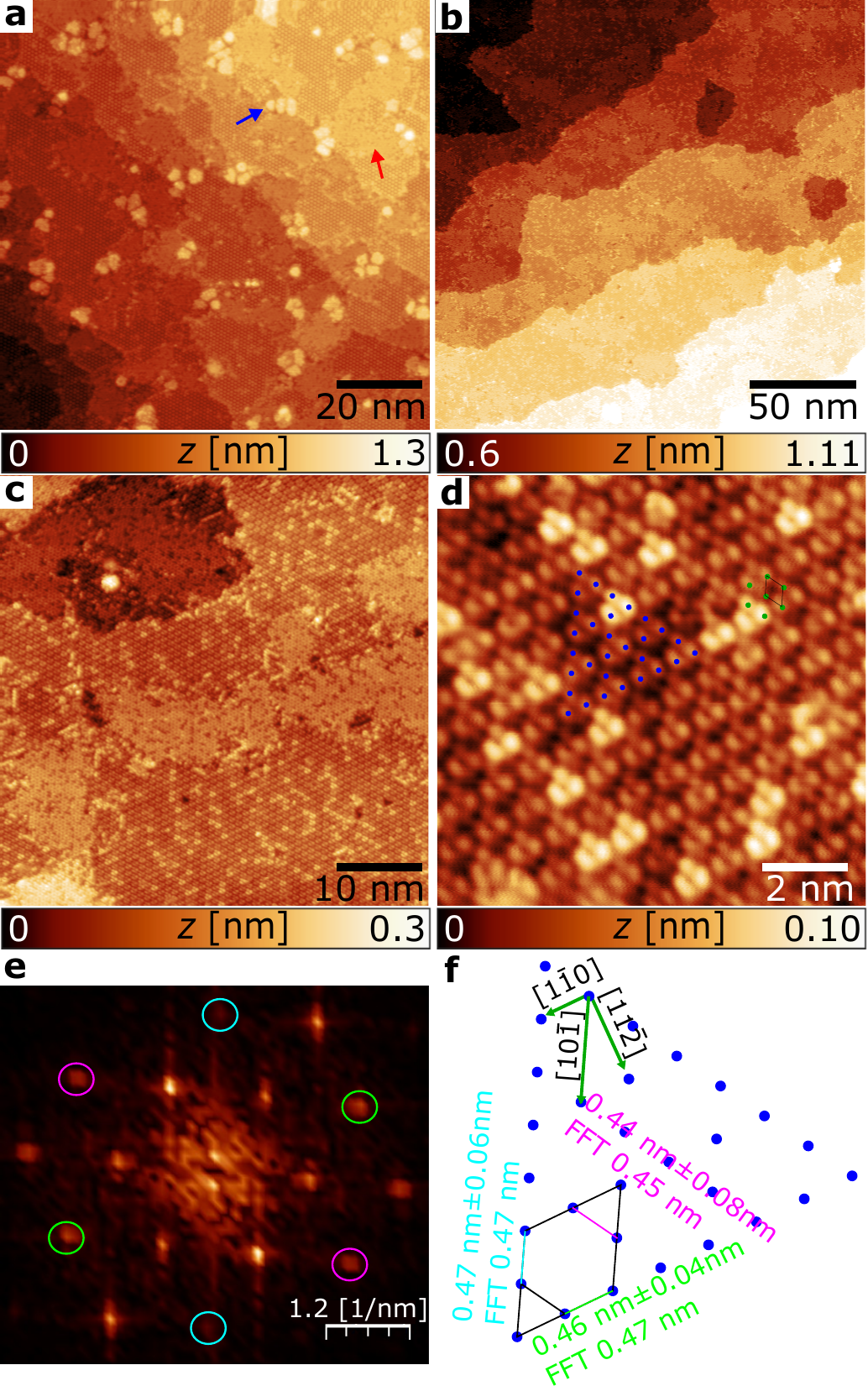}
	\caption{\label{fig:Nb111_3}
		STM images showing the Nb(111) surface after treatment with atomic hydrogen and atomic-resolution STM data of the Nb(111) surface. 
		(\textbf{a}) Upon further treatment of the surface by short flashes, the Nb(111) surface predominantly exhibits a hexagonal superlattice structure. In addition, a differently structured surface (as indicated by a red arrow) as well as some small clusters (blue arrow) are visible. (\textbf{b}) STM image showing extended terraces after additional short flashes. (\textbf{c}) Close-up view of (b) with atomic resolution. (\textbf{d}) STM image showing a reconstructed surface with blue dots marking the atomic positions of certain atoms to illustrate the reconstructed lattice structure, with the unit cell drawn in (\textbf{f}). In addition, one area is marked with green dots where no missing atom is visible within a hexagon. In (\textbf{e}) the Fourier transform of (\textbf{d}) is shown, with the corresponding distances color-coded and given in (\textbf{f}). Furthermore, the grid drawn in (\textbf{d}) is shown enlarged in (\textbf{f}) with distances as deduced from the STM data and FFT.}
\end{figure}

\begin{figure}	
	\includegraphics[scale=0.8]{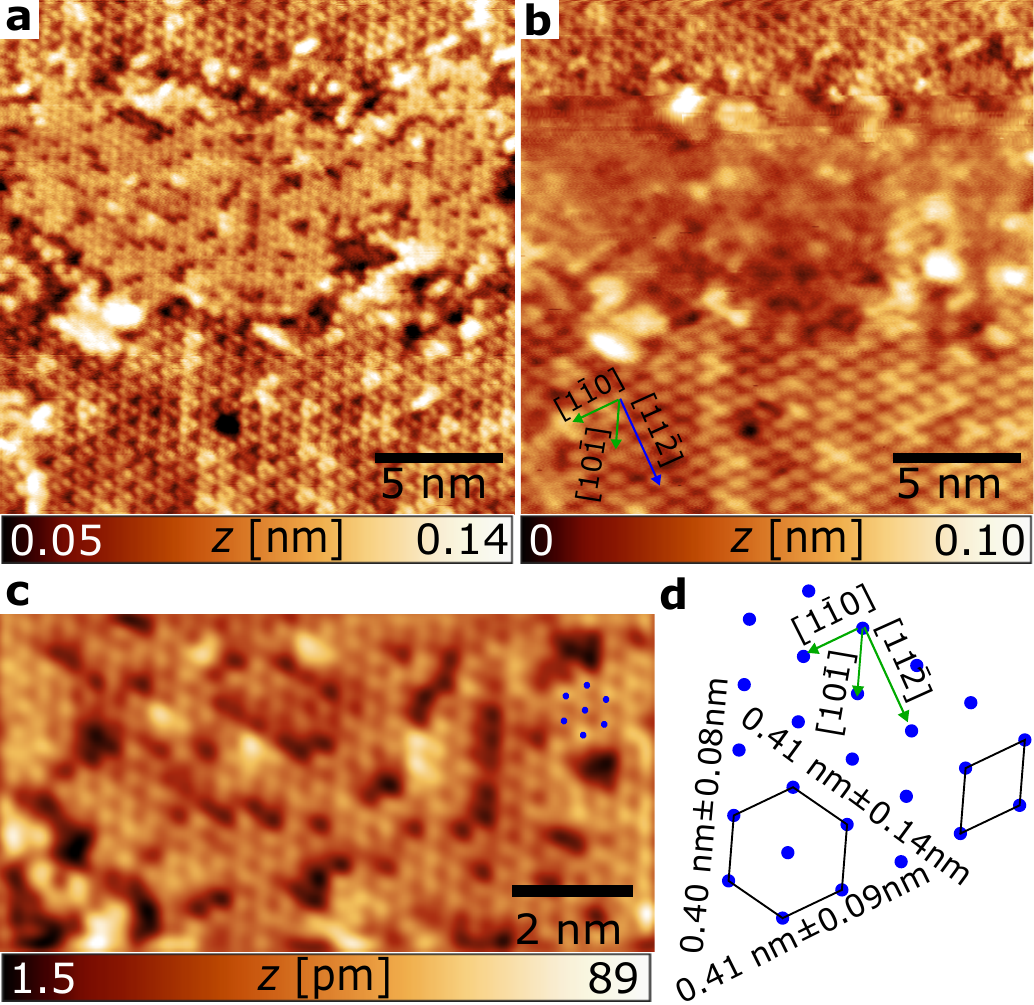}
	\caption{\label{fig:Nb111_4}
		STM images from the same location on the Nb(111) surface after hydrogen treatment and enlargement of the atomically resolved STM data of the densely packed hexagonal Nb(111) surface. 
		(\textbf{a}) Atomic-resolution STM data in contrast to (\textbf{b}) where no atomic resolution is observed for the lower part of the imaged area due to an STM tip change. The crystallographic directions are indicated by arrows. 
		(\textbf{c}) STM data showing the hexagonal lattice of clean Nb(111), together with atomic defects. (\textbf{d}) Schematic of the non-reconstructed Nb(111) lattice with atomic distances as deduced from the STM data (blues dots in (\textbf{c})).
		STM measurements parameters: (\textbf{a}) $V$~=~\SI{-500}{mV}, $I$~=~\SI{1}{nA}, (\textbf{b}) $V$~=~\SI{1}{V}, $I$~=~\SI{1}{nA} and (\textbf{c}) $V$~=~\SI{-100}{mV}, $I$~=~\SI{1}{nA} with $T$~=~\SI{25}{K}.}
\end{figure}

After concluding that the physical cleaning method was unsuccessful, a chemical approach was attempted. For this, the Nb(111) surface was exposed to atomic hydrogen followed by five short flashes (each with a duration of \SI{5}{s}) at $T~\approx$~\SI{1273}{K}, resulting in a change of the surface morphology as observed in Fig.~\ref{fig:Nb111_3} \textbf{a}: 
The surface clearly shows the underlying hexagonal superlattice structure which was already described in the context of Fig.~\ref{fig:Nb111_1}. In addition, the amount of elongated clusters decreased and higher clusters of about \SI{0.3}{nm}-\SI{0.4}{nm} (indicated with a blue arrow) are visible now. Also, other areas can be seen, which exhibit a denser structure than the hexagonal superlattice, with one area in the image marked with a red arrow for clarity. After five more flashes, the elevated regions are no longer present (see Fig.~\ref{fig:Nb111_3} \textbf{b} and \textbf{c}). In addition, the formation of large (more than \SI{50}{nm} wide) terraces is observed. 
Much more interesting, however, are the atomic-resolution images of the hexagonal superlattice, as presented in Fig.~\ref{fig:Nb111_3} \textbf{c}-\textbf{d}, Fig.~\ref{fig:Nb111_4} \textbf{a}, and partially seen in Fig.~\ref{fig:Nb111_4} \textbf{b}.
These atomically resolved STM data sets indicate that earlier STM images of the hexagonal superlattice structure did not represent the true atomic-scale structure of the Nb(111) surface. This becomes clear when Fig.~\ref{fig:Nb111_4} \textbf{a} and \textbf{b} are compared with each other. Both figures show STM images of exactly the same area of the sample surface. Fig.~\ref{fig:Nb111_4} \textbf{a} reveals clearly the atomic-scale structure of the surface. On the other hand, in Fig.~\ref{fig:Nb111_4} \textbf{b}, after about one-fifth of the image has been scanned (from the top), a change of the STM tip occurred, now revealing the familiar hexagonal superlattice with twice the lattice constant of Nb(111) (bottom of Fig.~\ref{fig:Nb111_4} \textbf{b}). A close-up view of the surface structure with atomic resolution is shown in Fig.~\ref{fig:Nb111_3} \textbf{d}.
The underlying atomic lattice of the surface reconstruction is also sketched, marked with blue dots. 
It is noticeable that for most hexagons, the atom in the center seems to be missing and thus this Nb(111) surface exhibits a Kagome-like lattice structure. However, areas where this is not the case (see atoms marked by green dots) can also be found in the STM image. Measurements in both real and reciprocal space (after Fourier transform Fig.~\ref{fig:Nb111_3} \textbf{e}) of the shortest atomic distances in  $[1\bar{1}0]$, $[10\bar1]$, and $[01\bar{1}]$ directions are shown in Fig.~\ref{fig:Nb111_3} \textbf{f}, agreeing very well with the theoretical value of the lattice constant of Nb(111), which is $a_\mathrm{Nb}$~=~\SI{0.467}{nm}. The typical apparent height of the atoms is approx.~\SI{20}{pm}. The unit cell of the Kagome-like structure is shown in Fig.~\ref{fig:Nb111_3} \textbf{f} (black). 
The Kagome-like lattice is qualitatively consistent with the theoretical work of Yang et al.~\cite{Yang2007}, who investigated melting mechanisms of Nb(111) planes. It has been shown by Yang et al.~that Nb(111) should exhibit an initial activation of subsurface atoms already at a temperature of $T\approx$~\SI{1250}{K}. Yang et al.~describe this process as premelting. In this process, the density of the first atomic layers is slightly reduced compared to the layers below.\\
Looking closely at the close-packed structure shown enlarged in Fig.~\ref{fig:Nb111_4} \textbf{c}, a hexagonal lattice can be identified, though with some atomic defects being present. The determination of the atomic distances from the STM image (Fig.~\ref{fig:Nb111_4} \textbf{c})) yields values that are in the range of the expected atomic lattice spacing of Nb(111) (Fig.~\ref{fig:Nb111_4} \textbf{d})). This result suggests that the hexagonal lattice shown in Fig.~\ref{fig:Nb111_4} \textbf{c} represents the unreconstructed Nb(111) surface, while pre-melting by heating leads to lower atomic density compared to the truncated (111) surface of a bcc crystal. \\

\section{Conclusion}
In this work, we have shown that it is very challenging to clean a contaminated Nb(111) surface in a reasonable time using standard sputtering and annealing techniques. The experimental STM data show that if the annealing temperature is too high ($T~\approx$~\SI{2463}{K} with an annealing time of 10 min), the Nb(111) surface shows facetting, leading to $\left\lbrace 211\right\rbrace$ pyramidal structures \cite{Madey1999}. Note that facetting can also occur at lower temperatures, depending on the duration of the annealing. An alternative treatment with hydrogen results in two different surface structures, namely reconstructed Nb(111) which exhibits a Kagome-like lattice and an atomic hexagonal lattice (with some vacancy sites). For both, a lattice spacing that is in the range of the theoretical value of the lattice constant of Nb(111) $a_\mathrm{Nb}$~=~\SI{0.467}{nm} can be found. These two surfaces most likely represent different stages of premelting, as theoretically predicted by Yang et al.~\cite{Yang2007}. In conclusion, a very high surface quality of Nb(111) has been achieved with wide terraces and a very low concentration of residual surface defects, which provides a sound basis for realizing well defined magnet - superconductor hybrid systems involving Nb(111) substrates.

\section*{Acknowledgments}
We thank J. Wiebe for fruitful discussions. We gratefully acknowledge funding by the European Union via the ERC Advanced Grant ADMIRE (No. 786020). M.B. acknowledges the Polish Ministry of Education and Science within Project No. 0512/SBAD/2220 and the National Science Center Poland via the Sonata Bis Project No. 2017/26/E/ST3/00140.


\providecommand{\latin}[1]{#1}
\makeatletter
\providecommand{\doi}
{\begingroup\let\do\@makeother\dospecials
	\catcode`\{=1 \catcode`\}=2 \doi@aux}
\providecommand{\doi@aux}[1]{\endgroup\texttt{#1}}
\makeatother
\providecommand*\mcitethebibliography{\thebibliography}
\csname @ifundefined\endcsname{endmcitethebibliography}
{\let\endmcitethebibliography\endthebibliography}{}


\end{document}


\newpage
\section{Supplementary Note 1\\ Analysis of the pyramid structure after intensive annealing}
\begin{figure}
	\includegraphics[width=0.8\textwidth]{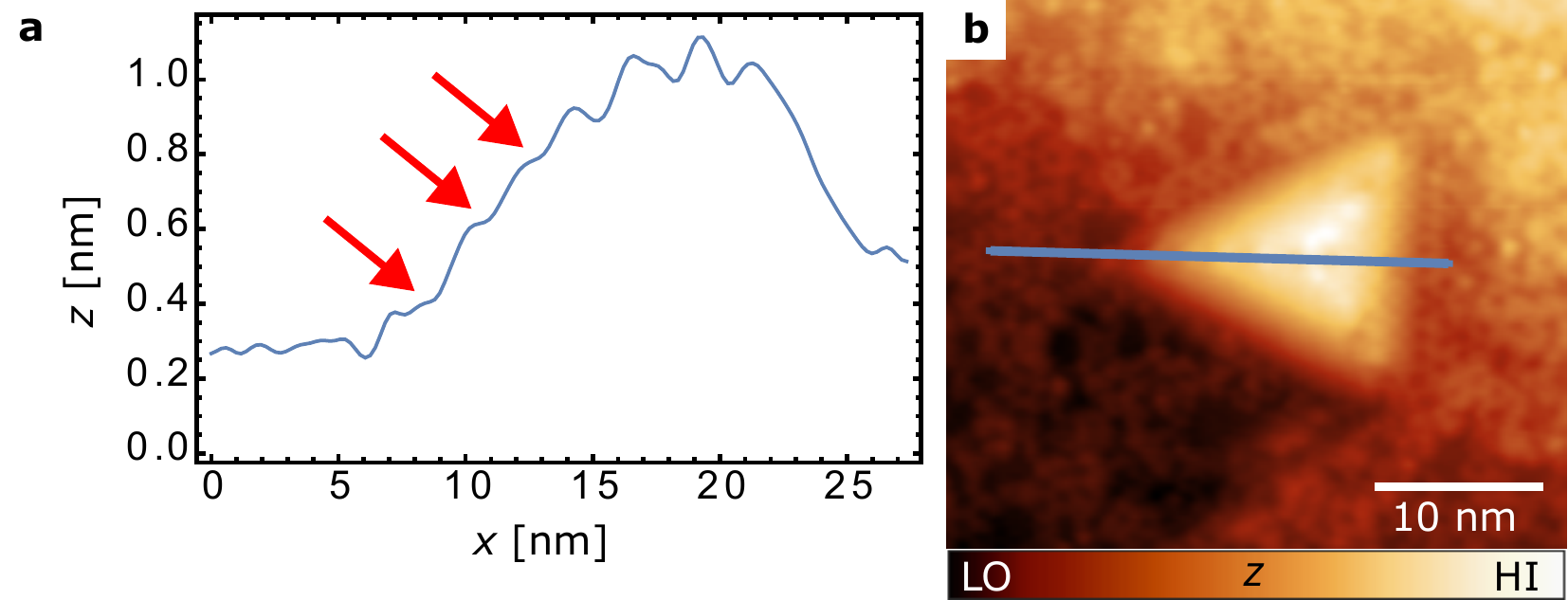}
	\caption{(\textbf{a}) Profile line across the pyramid structure in the STM image in panel (\textbf{b}). The red arrows indicate step edges that were clearly found and most likely support the hypothesis of a $\left\lbrace 211\right\rbrace$ pyramidal structure. Measurement parameters: $V$~=~\SI{2}{V}, $I$~=~\SI{50}{pA}, $T$~=~\SI{25}{K}.}
	\label{fig:S1}
\end{figure}

As already described in the main text, the surface exhibits facetting upon increased heating. For the analysis of the pyramid structure, Figure~\ref{fig:S1} (\textbf{a}) shows a profile line, which was taken from the STM image shown in Figure~\ref{fig:S1} (\textbf{b}), where step edges are clearly visible. With the given resolution, this most likely indicates a $\left\lbrace 211\right\rbrace$ pyramidal structure. For a $\left\lbrace 011\right\rbrace$ pyramidal structure, the step edges would not be as clearly visible because the atoms at the step edges are more densely packed than in a $\left\lbrace 211\right\rbrace$ pyramidal structure. See also Figure 2 \textbf{f} in the main text, where the step edges are indicated by the red arrows.

\newpage

\section{Supplementary Note 2\\ Analysis of the bright clusters after hydrogen treatment}
In Figure 3 (\textbf{d}) of the main text, some atoms, usually forming a three-atom cluster, appear very high compared to the neighboring atoms (in some cases two times higher). 
These three-atom clusters, which appear as bright features may still have oxygen attached. This assumption can be supported by looking at d$I$/d$V$ map in the Supplement Figure \ref{fig:S2} \textbf{b} taken at $V$~=~\SI{-100}{mV} with the corresponding topography presented in Figure \ref{fig:S2} \textbf{a}. The three-atom clusters that appear bright show strongly reduced d$I$/d$V$ values at this energy indicating a strongly reduced local DOS at this location. In contrast, the dense structure (in the upper part of the image) and the rest of the Kagome-like lattice clearly show a stronger differential tunneling conductance. This suggests that another elemental species is involved or attached to the bright three atoms, which most likely is oxygen.\\

\begin{figure}
	\includegraphics[width=0.8\textwidth]{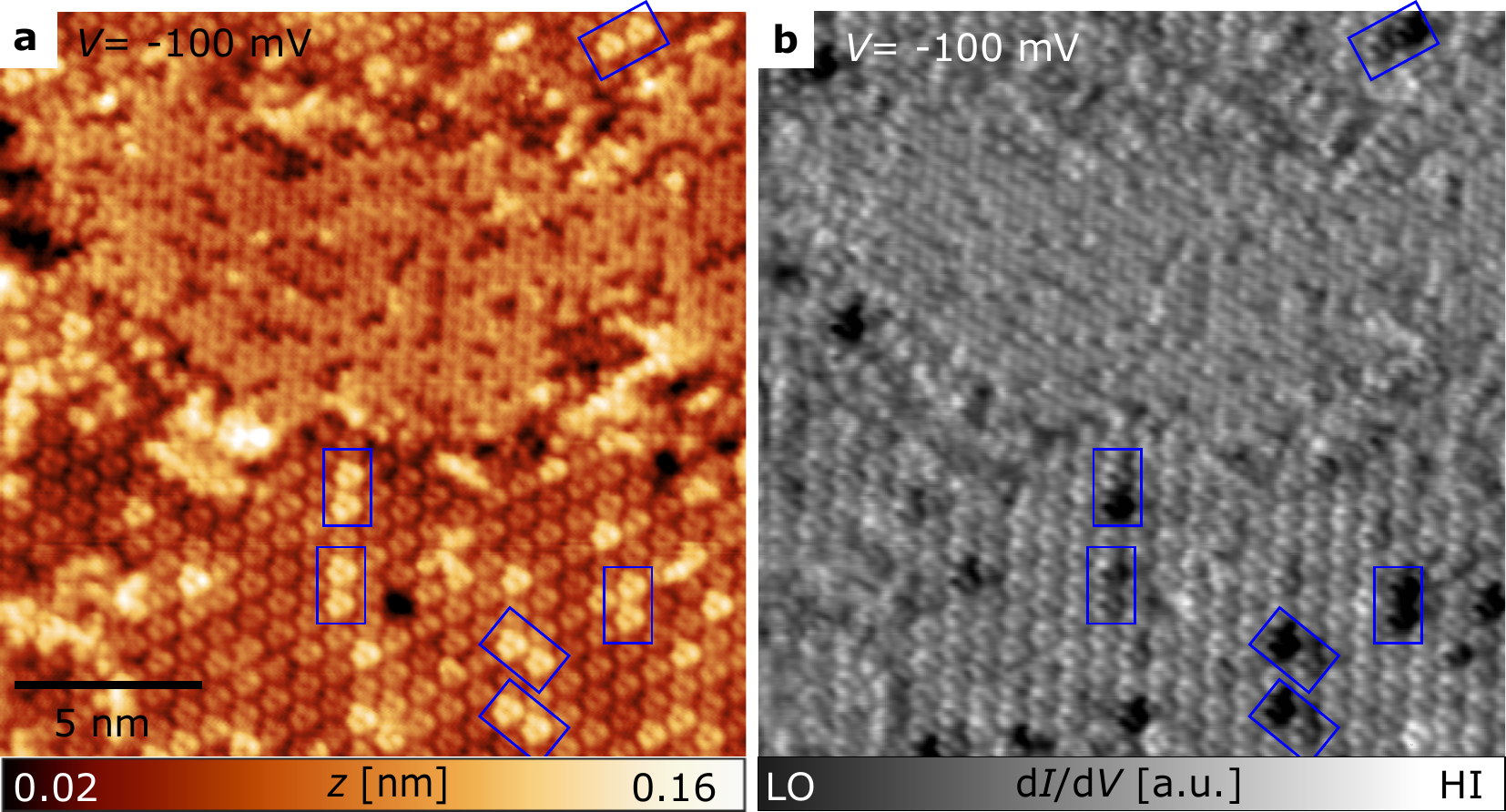}
	\caption{Atomically resolved STM data of the reconstructed Nb(111) surface. The Kagome-like lattice and close-packed structure are shown in (\textbf{a}) with the simultaneously acquired d$I$/d$V$ map in (\textbf{b}), where the bright three-atom clusters (framed in blue) show strongly reduced d$I$/d$V$ values at $V$~=~\SI{-100}{mV}, unlike the non-bright neighboring atoms. STM measurement parameters for \textbf{a} and \textbf{b}: $V$~=~\SI{-100}{mV}, $V_\mathrm{mod}$~=~\SI{10}{mV}, $I$~=~\SI{1}{nA}, $T$~=~\SI{25}{K}.}
	\label{fig:S2}
\end{figure}

\newpage
